\documentclass[preprintnumbers,amsmath,amssymb,floatfix,11pt,prd,onecolumn,showpacs,
superscriptaddress,nofootinbib]{revtex4}
\usepackage{graphicx}
\usepackage{epsfig}
\usepackage{bm}
\usepackage{amsfonts}

\def\ben{\begin{equation}}
\def\een{\end{equation}}

 \def\bd{\begin{document}} \def\ed{\end{document}}
\def\ds{\documentstyle} \let\fr=\frac \let\bl=\bigl \let\br=\bigr
\let\Br=\Bigr \let\Bl=\Bigl
\let\bm=\bibitem
\let\na=\nabla
\let\pa=\partial \let\ov=\overline
\newcommand{\be}{\begin{equation}}
\newcommand{\ee}{\end{equation}}
\def\ba{\begin{array}}
\def\ea{\end{array}}
\def\ft#1#2{{\textstyle{\frac{\scriptstyle #1}{\scriptstyle #2} } }}
\def\fft#1#2{{\frac{#1}{#2}}}
\def\del{\partial}
\def\vp{\varphi}
\def\sst#1{{\scriptscriptstyle #1}}
\def\oneone{\rlap 1\mkern4mu{\rm l}}
\def\td{\tilde}
\def\wtd{\widetilde}
\def\ie{{\it i.e.\ }}
\def\dalemb#1#2{{\vbox{\hrule height .#2pt
        \hbox{\vrule width.#2pt height#1pt \kern#1pt
                \vrule width.#2pt}
        \hrule height.#2pt}}}
\def\square{\mathord{\dalemb{6.8}{7}\hbox{\hskip1pt}}}
\newcommand{\ho}[1]{$\, ^{#1}$}
\newcommand{\hoch}[1]{$\, ^{#1}$}
\newcommand{\bea}{\setlength\arraycolsep{2pt} \begin{eqnarray}}
\newcommand{\eea}{\end{eqnarray}}
\newcommand{\ra}{\rightarrow}
\newcommand{\lra}{\longrightarrow}
\newcommand{\Lra}{\Leftrightarrow}
\newcommand{\bp}{\tilde \beta^\prime}
\newcommand{\tr}{{\rm tr} }
\newcommand{\Tr}{{\rm Tr} }
\def\0{{\sst{(0)}}}
\def\1{{\sst{(1)}}}
\def\2{{\sst{(2)}}}
\def\3{{\sst{(3)}}}
\def\4{{\sst{(4)}}}
\def\5{{\sst{(5)}}}
\def\6{{\sst{(6)}}}
\def\7{{\sst{(7)}}}
\def\8{{\sst{(8)}}}
\def\m{{\sst{(m)}}}
\def\n{{\sst{(n)}}}
\def\cA{{{\cal A}}}
\def\cB{{{\cal B}}}
\def\cF{{{\cal F}}}
\def\cG{{{\cal G}}}
\def\cH{{{\cal H}}}
\def\tV{\widetilde V}
\def\tW{\widetilde W}
\def\tH{\widetilde H}
\def\tE{\widetilde E}
\def\tF{\widetilde F}
\def\tA{\widetilde A}
\def\im{{{\rm i}}}
\def\tY{{{\wtd Y}}}
\def\ep{{\epsilon}}
\def\vep{{\varepsilon}}
\def\bD{{{\bar D}}}
\def\R{{{\mathbb R}}}
\def\C{{{\mathbb C}}}
\def\H{{{\mathbb H}}}
\def\CP{{{\mathbb C}{\mathbb P}}}
\def\RP{{{\mathbb R}{\mathbb P}}}
\def\Z{{{\mathbb Z}}}
\def\bA{{{\mathbb A}}}
\def\bB{{{\mathbb B}}}
\def\bC{{{\mathbb C}}}
\def\bD{{{\mathbb D}}}
\def\bE{{{\mathbb E}}}
\def\bZ{{{\mathbb Z}}}
\def\Re{{{\frak{Re}}}}
\def\Im{{{\frak{Im}}}}
\def\cosec{{\,\hbox{cosec}\,}}
\def\Gm{{\Gamma_{\!\! -}}}
\def\Gp{{\Gamma_{\!\! +}}}
\def\stan{{standard }}
\def\nonstan{{supernumerary }}
\def\p{{\partial}}
\def\kdel#1{{\fft{\del}{\del#1}}}

\def\bog{{Bogomolny}}
\def\om{{\omega}}

\newcommand{\nnr}{\nonumber \\}
\newcommand{\pd}{\partial}
\newcommand{\ud}{\textrm{d}}
\newcommand{\dTH}{T^{\prime \, 0}_\textrm{H}}
\newcommand{\dOi}{\Omega^{\prime \, 0}_i}
\newcommand{\bx}{{\bf x}}
\begin{document}

\title{Entropy-corrected new agegraphic
 dark energy in Ho\v{r}ava-Lifshitz cosmology}

\author{K. Karami}
\email{KKarami@uok.ac.ir} \affiliation{Department of Physics,
University of Kurdistan, Pasdaran St., Sanandaj,
Iran}\affiliation{Research Institute for Astronomy $\&$ Astrophysics
of Maragha (RIAAM), Maragha, Iran}
\author{M. Jamil}
\affiliation{Center for Advanced Mathematics and Physics, National
University of Sciences and Technology, Islamabad, Pakistan}
\author{M. Roos}
\affiliation{Department of Physics, University of Helsinki,
Helsinki, Finland}
\author{S. Ghaffari}
\affiliation{Department of Physics, University of Kurdistan,
Pasdaran St., Sanandaj, Iran}
\author{A. Abdolmaleki}
\affiliation{Department of Physics, University of Kurdistan,
Pasdaran St., Sanandaj, Iran}

\begin{abstract}
We study the entropy-corrected version of the new agegraphic dark
energy (NADE) model and dark matter in a spatially non-flat Universe
and in the framework of Ho\v{r}ava-Lifshitz cosmology. For the two
cases containing noninteracting and interacting entropy-corrected
NADE (ECNADE) models, we derive the exact differential equation that
determines the evolution of the ECNADE density parameter. Also the
deceleration parameter is obtained. Furthermore, using a
parametrization of the equation of state parameter of the ECNADE
model as $\omega_{\Lambda}(z)=\omega_0+\omega_1 z$, we obtain both
$\omega_0$ and $\omega_1$. We find that in the presence of
interaction, the equation of state parameter $\omega_0$ of this
model can cross the phantom divide line which is compatible with the
observation.
\end{abstract}
\maketitle

\section{Introduction}
Astronomical observations indicate that our Universe is in a phase
of accelerated expansion (Perlmutter et al. 1999; Bennett et al.
2003; Tegmark et al. 2004; Allen et al. 2004). One explanation for
the cosmic acceleration is the dark energy (DE), an exotic energy
with negative pressure. The dynamical nature of DE, at least in an
effective level, can originate from various fields, although a
complete description requires a deeper understanding of the
underlying theory of quantum gravity. Nevertheless, physicists can
still make some attempts to probe the nature of DE according to some
basic quantum gravitational principles. Two examples of such a
paradigm are the holographic DE (HDE) and the agegraphic DE (ADE)
models which have originated from quantum gravity and possess some
of its significant features. The former, that arose a lot of
enthusiasm recently (Cohen et al. 1999; Hsu 2004; Li 2004; Huang and
Li 2004; Jamil et al. 2009a, 2009b; Jamil and Farooq 2010; Setare
and Jamil 2010; Wang et al. 2005a, 2005b, 2008; Sheykhi 2010a;
Karami 2010a, 2010b; Karami and Abdolmaleki 2010a, 2010b; Karami and
Fehri 2010a, 2010b), is motivated from the holographic hypothesis
('t Hooft 1993; Susskind 1995) and has been tested and constrained
by various astronomical observations (Feng et al. 2005; Zhang and Wu
2005, 2007). The latter originated from the uncertainty relation of
quantum mechanics together with the gravitational effect of general
relativity (GR). The ADE model assumes that the observed DE effect
comes from spacetime and matter field fluctuations in the Universe
(Sasakura 1999; Cai 2007; Wei and Cai 2008a, 2009). Following the
line of quantum fluctuations of spacetime, Karolyhazy (Karolyhazy
1966; Karolyhazy et al. 1982, 1986) proposed that the distance $t$
in Minkowski spacetime cannot be known to a better accuracy than
$\delta t=\varepsilon t_{P}^{2/3}t^{1/3}$, where $\varepsilon$ is a
dimensionless constant of order unity and $t_P$ is the reduced
Planck time. Based on the Karolyhazy relation, Maziashvili proposed
that the energy density of metric fluctuations of Minkowski
spacetime is given by (Maziashvili 2007a, 2007b)
\begin{equation}\label{2}
\rho_{\Lambda}\sim\frac{1}{t_{P}^{2}t^{2}}\sim\frac{M_{P}^{2}}{t^{2}},
 \end{equation}
 where $M_P$ is the reduced Planck mass $M_P^{-2}=8\pi G$. Since in the original ADE model the
age of the Universe is chosen as the length measure, instead of the
horizon distance, the causality problem in the HDE is avoided (Cai
2007). The original ADE model had some difficulties. In particular,
it cannot justify the matter-dominated era (Cai 2007). This
motivated Wei and Cai (2008a) to propose the new ADE model, while
the time scale is chosen to be the conformal time instead of the age
of the Universe. The NADE density is given by (Wei and Cai 2008a)
\begin{equation}
\rho_{\Lambda}=\frac{3{n}^2M_P^2}{\eta^2},\label{NADE}
\end{equation}
where 3$n^2$ is the numerical factor and $\eta$ is the conformal
time and defined as
\begin{equation}
\eta=\int\frac{{\rm d}t}{a}=\int_0^a\frac{{\rm
d}a}{Ha^2}.\label{eta}
\end{equation}
The ADE models have been examined and constrained by various
astronomical observations (Kim et al. 2008a, 2008b; Wu et al. 2008;
Zhang et al. 2008; Neupane 2009; Sheykhi 2009a, 2009b, 2010b; Karami
et al. 2010; Karami and Abdolmaleki 2011; Wei and Cai 2008b).
Besides, in the loop quantum gravity, the entropy-area relation can
be modified due to the thermal equilibrium fluctuations and quantum
fluctuations (Banerjee et al. 2008; Banerjee and Majhi 2008a, 2008b,
2009; Banerjee and Modak 2009; Majhi 2009, 2010; Modak 2009). The
corrected entropy takes the form
\begin{equation}\label{S}
S=\frac{A}{4G}+\tilde{\alpha} \ln {\frac{A}{4G}}+\tilde{\beta},
\end{equation}
where $\tilde{\alpha}$ and $\tilde{\beta}$ are dimensionless
constants of order unity. Taking the corrected entropy-area relation
(\ref{S}) into account, the NADE density will be modified as well.
Motivated by the corrected entropy-area relation (\ref{S}), the
ECNADE density is given by (Wei 2009; Karami and Sorouri 2010;
Karami et al. 2011)
\begin{eqnarray}
\rho_{\Lambda} = \frac{3n^2{M_P^2}}{\eta^2} +
\frac{\alpha}{{\eta}^4}\ln{({M_P^2}{\eta}^2)} +
\frac{\beta}{\eta^4},\label{density-nade}
\end{eqnarray}
where $\alpha$ and $\beta$ are dimensionless constants of order
unity.

Recently a power-counting renormalizable UV complete theory of
gravity was proposed by Ho\v{r}ava (Ho\v{r}ava 2009a, 2009b).
 This theory is not Lorentz invariant (except in the
infrared limit), it is power-counting renormalizable and obeys
anisotropic scaling or Lifshitz scaling. Than the time coordinate
and the 3 spatial coordinates have to be treated separately, the
theory is non-relativistic, the speed of light diverges in the
ultraviolet limit, and test particles do not follow geodesics. In
consequence causal structures are different from that in General
Relativity (Greenwald 2011) and entropy cannot be defined. Quantum
gravity models based on an anisotropic scaling of the space and time
dimensions have recently attracted significant attention (Visser
2009; Pal 2009). In particular, Ho\v{r}ava-Lifshitz point gravity
might not have desirable features, but in its original incarnation
one is forced to accept a non-zero cosmological constant of the
wrong sign to be compatible with observations (Nastase 2009). There
are four different versions of this theory: with (or without)
projectability condition and with (or without) detailed balance. At
a first look it seems that this non-relativistic model for quantum
gravity has a well defined IR limit and it reduces to GR. But as it
was first indicated by Mukohyama (Mukohyama 2009, 2010; Saridakis
2010), Ho\v{r}ava-Lifshitz theory mimics GR plus dark matter (DM).
This theory has a scale invariant power spectrum which describes
inflation. For reviews on the scenario where the cosmological
evolution is governed by Ho\v{r}ava-Lifshitz gravity see (Mukohyama
2009, 2010; Saridakis 2010; Wang and Wu 2009; Majhi and Samanta
2010; Wang 2011; Khatua et al. 2011).

Here, our aim is to investigate the ECNADE model in
Ho\v{r}ava-Lifshitz cosmology. To do this in section 2, we first
review the scenario of Ho\v{r}ava-Lifshitz gravity. In section 3, we
consider a spatially non-flat Friedmann-Robertson-Walker (FRW)
Universe containing ECNADE and DM in the framework of
Ho\v{r}ava-Lifshitz gravity. We obtain the evolution of
dimensionless energy density, deceleration parameter and equation of
state of ECNADE with interaction/non-interaction. Section 4 is
devoted to conclusions.

\section{Ho\v{r}ava-Lifshitz gravity}
Under the projectability condition, the full metric in the (3+1)
dimensional Arnowitt-Deser-Misner formalism is written as (Calcagni
2009; Kiritsis and Kofinas 2009)
\begin{eqnarray}
ds^2 = - N^2 dt^2 + g_{ij} (dx^i + N^i dt )( dx^j + N^j dt ),
\end{eqnarray}
where the dynamical variables $N$, $N_i$ and $g_{ij}$ are the lapse
function, shift vector and 3-dimensional metric, respectively. Note
that the projectability condition restricts the lapse function $N$
to be space-independent, while the shift vector $N^i$ and the
3-dimensional metric $g_{ij}$ still depend on both time and space.

Under the detailed balance condition, the gravitational action of
Ho\v{r}ava-Lifshitz (HL) gravity is given by (Ho\v{r}ava 2009a,
2009b)
\begin{eqnarray}
 S_g &=&  \int dt d^3x \sqrt{g} N \Big[
\frac{2}{\kappa^2} (K_{ij}K^{ij} - \lambda K^2)\nonumber
\\&& +\frac{\kappa^2}{2
w^4} C_{ij}C^{ij} -\frac{\kappa^2 \mu}{2 w^2}
\frac{\epsilon^{ijk}}{\sqrt{g}} R_{il} \nabla_j R^l_k \nonumber
\\&&+
\frac{\kappa^2 \mu^2}{8} R_{ij} R^{ij}
     \Big. \nonumber \\&&
+\Big.    \frac{\kappa^2 \mu^2}{8( 3 \lambda-1)} \Big( \frac{1 - 4
\lambda}{4} R^2 + \Lambda  R - 3 \Lambda ^2 \Big) \Big],
\label{acct}
\end{eqnarray}
where
\begin{eqnarray}
K_{ij} = \frac{1}{2N} \Big( {\dot{g}_{ij}} - \nabla_i N_j - \nabla_j
N_i \Big),
\end{eqnarray}
is the extrinsic curvature and
\begin{eqnarray} C^{ij} \, = \, \frac{\epsilon^{ijk}}{\sqrt{g}} \nabla_k
\bigl( R^j_i - \frac{1}{4} R \delta^j_i \bigr),
\end{eqnarray}
is the Cotton tensor. Also $\epsilon^{ijk}$ is the totally
antisymmetric unit tensor, $\lambda$ is a dimensionless constant and
$\Lambda $ is a positive constant which as usual is related to the
cosmological constant in the IR limit. The variables $\kappa$, $w$
and $\mu$ are constants with mass dimensions $-1$, $0$ and $1$,
respectively. Note that the detailed balance condition restricts the
form of a general potential in a ($d$ + 1)-dimensional Lorentz
action to a specific form that can be expressed in terms of a
$d$-dimensional action of a relativistic theory with Euclidean
signature, whereby the number of independent couplings is
considerably limited.

In particular, Eq.(7) suffers several problems, including
instability, inconsistency and strong coupling problems, for detail,
see (Mukohyama,2010). To overcome these problems, one way is to
provoke the Vainshtein mechanism, as first done by Mukohyama for
spherical spacetimes (Mukohyama,2010) and by Wang and Wu in the
cosmological setting (Wang \& Wu,2011). Such considerations were
further carried out by using the so-called gradient expansion method
(Izumi \& Mukohyama,2011;Gumrukcuoglu et al,2011). Another very
promising approach is to introduce an extra $U(1)$ symmetry, as
first done by (Horava \& Melby-Thompson,2010)  with $\lambda =1$,
and later generalized to the case with any $\lambda$ by (da
Silva,2011). These studies were further generalized to the case
without the projectability condition (Zhu et al,2011a; Zhu et
al,2011b). In both cases (with/without the projectability
condition), due to the $U(1)$ symmetry, the spin-0 gravitons are
eliminated, and all the problems related to them, such as
instability, inconsistency and strong coupling problems, are
resolved.

To include the matter component we add a cosmological stress-energy
tensor to the gravitational field equations, by imposing a condition
to recover the general relativistic formalism in the low-energy
limit (Carloni et al. 2009; Leon and Saridakis 2009; Sotiriou et al.
2009; Chaichian et al. 2010; Dutta and Saridakis 2010). For this,
the energy density $\rho_{\rm m}$ and pressure $p_{\rm m}$ satisfy
the energy conservation equation
\begin{eqnarray}\label{rhodotfluid}
\dot{\rho}_{\rm m}+3H(\rho_{\rm m}+p_{\rm m})=0.
\end{eqnarray}
In the cosmological context, we use a FRW metric as
\begin{eqnarray}
N=1,~~g_{ij}=a^2(t)\gamma_{ij},~~N^i=0,
\end{eqnarray}
with
\begin{eqnarray}
\gamma_{ij}dx^idx^j=\frac{dr^2}{1- k r^2}+r^2d\Omega_2^2,
\end{eqnarray}
where $ k=-1,0,+1$ refer  to spatially open, flat, and closed
Universe respectively.

On varying the action with respect to the metric components $N$ and
$g_{ij}$, one can obtain the modified Friedmann equations in the
framework of HL gravity as
\begin{eqnarray}\label{Fr1fluid}
H^2 &=& \frac{\kappa^2}{6(3\lambda-1)} \rho_{\rm m}
+\frac{\kappa^2}{6(3\lambda-1)}\Big[ \frac{3\kappa^2\mu^2
k^2}{8(3\lambda-1)a^4}\nonumber \\ &&+\frac{3\kappa^2\mu^2\Lambda
^2}{8(3\lambda-1)}
 \Big]-\frac{\kappa^4\mu^2\Lambda  k}{8(3\lambda-1)^2a^2},
\end{eqnarray}
\begin{eqnarray}\label{Fr2fluid}
\dot{H}+\frac{3}{2}H^2 &=& -\frac{\kappa^2}{4(3\lambda-1)} p_{\rm m}
\nonumber \\
&&-\frac{\kappa^2}{4(3\lambda-1)}\Big[\frac{\kappa^2\mu^2
k^2}{8(3\lambda-1)a^4} -\frac{3\kappa^2\mu^2\Lambda
^2}{8(3\lambda-1)}
 \Big]\nonumber \\&&-\frac{\kappa^4\mu^2\Lambda  k}{16(3\lambda-1)^2a^2},
\end{eqnarray}
where $H=\frac{\dot a}{a}$ is the Hubble parameter.

Noticing the form of the above Friedmann equations, we can define
for DE
\begin{equation}\label{rhoDE}
\rho_\Lambda\equiv \frac{3\kappa^2\mu^2 k^2}{8(3\lambda-1)a^4}
+\frac{3\kappa^2\mu^2\Lambda ^2}{8(3\lambda-1)},
\end{equation}
\begin{equation}
\label{pDE} p_{\Lambda}\equiv \frac{\kappa^2\mu^2
k^2}{8(3\lambda-1)a^4} -\frac{3\kappa^2\mu^2\Lambda
^2}{8(3\lambda-1)}.
\end{equation}
The first term on the right hand side proportional to $a^{-4}$ is
effectively the ``dark radiation term'', present in HL cosmology
(Calcagni 2009; Kiritsis and Kofinas 2009), while the second term is
referred as an explicit cosmological constant. Hence, in expressions
(\ref{rhoDE}) and (\ref{pDE}) we defined the energy density and
pressure for the effective DE, which include the aforementioned
contributions. Finally, note that using (\ref{rhoDE}) and
(\ref{pDE}) it is easy to show that these satisfy the following
expression
\begin{eqnarray}
\label{DEevol} \dot{\rho}_\Lambda+3H(\rho_\Lambda+p_\Lambda)=0.
\end{eqnarray}
Finally in order for these expressions to match with the standard
Friedmann equations ($c=1$) we set (Calcagni 2009; Kiritsis and
Kofinas 2009)
\begin{eqnarray}
G_{\rm
c}=\frac{\kappa^2}{16\pi(3\lambda-1)},\label{simpleconstants0a}
\end{eqnarray}
\begin{eqnarray}
\frac{\kappa^4\mu^2\Lambda}{8(3\lambda-1)^2}=1,
\label{simpleconstants0}
\end{eqnarray}
where $G_{\rm c}$ is the ``cosmological'' Newton's constant. Note
that in gravitational theories with the violation of Lorentz
invariance (like HL gravity) the ``gravitational'' Newton's constant
$G_{\rm g}$, which is present in the gravitational action, differs
with the ``cosmological'' Newton's constant $G_{\rm c}$, which is
present in Friedmann equations, unless Lorentz invariance is
restored (Carroll and Lim 2004). For the sake of completeness we
write
\begin{eqnarray}
G_{\rm g}=\frac{\kappa^2}{32\pi}\label{Ggrav}.
\end{eqnarray}
Note that in the IR limit ($\lambda=1$), where Lorentz invariance is
restored, $G_{\rm c}$ and $G_{\rm g}$ are the same.

Further we can rewrite the modified Friedmann equations
(\ref{Fr1fluid}) and (\ref{Fr2fluid}) in the usual form as
\begin{equation}
\label{eqfr} H^2+\frac{k}{a^2} = \frac{8\pi G_{\rm c}}{3}(\rho_{\rm
m}+\rho_\Lambda),
\end{equation}
\begin{equation}
\label{Fr2b} \dot{H}+\frac{3}{2}H^2+\frac{k}{2a^2} = - 4\pi G_{\rm
c}(p_{\rm m}+p_\Lambda).
\end{equation}

\section{ECNADE in HL cosmology}
Here we would like to investigate the ECNADE in HL theory. To do
this we consider a spatially non-flat FRW Universe containing the
ECNADE and DM. Let us define the dimensionless energy densities as
\begin{eqnarray}
\Omega_{\rm m}&=&\frac{\rho_{\rm m}}{\rho_{\rm cr}}=\frac{8\pi {\rm
G_{c}}}{3H^2}\rho_{\rm m},\nonumber\\ \Omega_{\rm
\Lambda}&=&\frac{\rho_{\Lambda}}{\rho_{\rm cr}}=\frac{8\pi {\rm
G_{c}}}{3H^2}\rho_{\Lambda},\nonumber\\
\Omega_{k}&=&-\frac{k}{a^2H^2}, \label{eqomega}
\end{eqnarray}
then, the first Friedmann equation (\ref{Fr1fluid}) yields
\begin{equation}
1-\Omega_{k}=\Omega_{\Lambda}+\Omega_{\rm m}.\label{eq10}
\end{equation}
From definition $\rho_{\Lambda}=\frac{3H^2}{8\pi G_{\rm
c}}\Omega_{\Lambda}$, we get
\begin{eqnarray}
\Omega_{\Lambda} =
\frac{n^2}{H^2\eta^2}\gamma_{n},\label{density-nade-omega}
\end{eqnarray}
where
\begin{eqnarray}
\label{gamma-parameter1} \gamma_n = \frac{G_{\rm c}}{G_{\rm g}} +
\frac{8\pi G_{\rm c}}{3n^2\eta^2}\Big[\alpha\ln{({M_P^2}{\eta}^2)}
+\beta\Big],
\end{eqnarray}
and $M_P^{-2}=8\pi G_{\rm g}$. Taking time derivative of Eq.
(\ref{density-nade}) and using $\dot{\eta}=1/a$, one can get
\begin{equation}
\dot{\rho}_{\Lambda}=\Big(\frac{2}{a\eta}\Big)\Big[-2\rho_{\Lambda}+\frac{3n^2M_P^2}{\eta^2}+\frac{\alpha}{\eta^4}\Big].\label{rhodot}
\end{equation}
Taking time derivative of $\Omega_{\Lambda}=\frac{8\pi G_{\rm
c}}{3H^2}\rho_{\Lambda}$ and using Eqs. (\ref{density-nade-omega}),
(\ref{rhodot}), $\dot{\eta}=1/a$ and
${\Omega^{\prime}_{\Lambda}}=\dot{\Omega_{\Lambda}}/H$, one can
obtain the equation of motion for $\Omega_{\Lambda}$ as
\begin{eqnarray}
\label{omegaD-eq-motion1} {\Omega^{\prime}_{\Lambda}}& =&
-2\Omega_{\Lambda}
\Big[\frac{\dot{H}}{H^2}+\frac{1}{na\gamma_n}\Big({\frac{\Omega_{\Lambda}}{\gamma_n}}\Big)
^{1/2}\Big(2\gamma_n - \frac{G_{\rm c}}{G_{\rm
g}}\nonumber\\&&-\frac{8\pi G_{\rm c}\alpha
H^2}{3n^4}\frac{\Omega_{\Lambda}}{\gamma_n}\Big)\Big].
\end{eqnarray}
Here, prime denotes the derivative with respect to $x=\ln{a}$.
Taking derivative of $\Omega_k=-k/(a^2H^2)$ with respect to $x=\ln
a$, one gets
\begin{equation}
{\Omega^{\prime}_{k}}=-2\Omega_{k}\Big(1+\frac{\dot{H}}{H^2}\Big).\label{omegak-eq-motion1}
\end{equation}
\subsection{Noninteracting case}
Consider the FRW Universe filled with ECNADE and pressureless DM
which evolves according to their conservation laws
\begin{equation}
\dot{\rho}_{\Lambda}+3H(1+\omega_{\Lambda})\rho_{\Lambda}=0,\label{eqpol}
\end{equation}
\begin{equation}
\dot{\rho}_{\rm m}+3H\rho_{\rm m}=0,\label{eqCDM}
\end{equation}
where $\omega_{\Lambda}=p_{\Lambda}/\rho_{\Lambda}$ is the equation
of state (EoS) parameter of the ECNADE model.

Taking time derivative of the first Friedmann equation
(\ref{Fr1fluid}) and using Eqs. (\ref{eqomega}), (\ref{eq10}),
(\ref{density-nade-omega}), (\ref{gamma-parameter1}), (\ref{rhodot})
and (\ref{eqCDM}), one can get
\begin{eqnarray}
\frac{\dot{H}}{H^2} =\frac{1}{2}\Big[\Omega_k-3(1 -
\Omega_{\Lambda})\Big]
~~~~~~~~~~~~~~~~~~~~~~~~~~~~~~~\nonumber\\-\frac{1}{na}\Big({\frac{\Omega_{\Lambda}}{\gamma_n}}\Big)^{3/2}\Big(2\gamma_n
- \frac{G_{\rm c}}{G_{\rm g}}-\frac{8\pi G_{\rm c}\alpha
H^2}{3n^4}\frac{\Omega_{\Lambda}}{\gamma_n}\Big).\label{H-dot-to-H2}
\end{eqnarray}
Substituting this into Eq. (\ref{omegaD-eq-motion1}), one obtains
\begin{eqnarray}
{\Omega^{\prime}_{\Lambda}}= \Omega_{\Lambda} \Big[3(1 -
\Omega_{\Lambda})-
\Omega_k+\frac{2}{na}\Big({\frac{\Omega_{\Lambda}}{\gamma_n}}\Big)^{1/2}
~~~~~~~~~~~\nonumber\\\times\Big(\frac{\Omega_{\Lambda}-1}{\gamma_n}\Big)\Big(2\gamma_n
-\frac{G_{\rm c}}{G_{\rm g}}-\frac{8\pi G_{\rm c}\alpha
H^2}{3{n^4}}\frac{\Omega_{\Lambda}}{\gamma_n}\Big)\Big].\label{omegaD-eq-motion2}
\end{eqnarray}
Putting Eq. (\ref{H-dot-to-H2}) into (\ref{omegak-eq-motion1})
reduces to
\begin{eqnarray}
{\Omega^{\prime}_{k}}=\Omega_{k}\Big[(1 -\Omega_k)-3\Omega_{\Lambda}
~~~~~~~~~~~~~~~~~~~~~~~~~~~~~~~~~\nonumber\\+\frac{2}{na}\Big({\frac{\Omega_{\Lambda}}{\gamma_n}}\Big)^{3/2}\Big(2\gamma_n
-\frac{G_{\rm c}}{G_{\rm g}}-\frac{8\pi G_{\rm c}\alpha
H^2}{3{n^4}}\frac{\Omega_{\Lambda}}{\gamma_n}\Big)\Big].\label{omega-prime-k}
\end{eqnarray}
The deceleration parameter is given by
\begin{equation}
q=-\Big(1+\frac{\dot{H}}{H^2}\Big).\label{q1}
\end{equation}
Replacing the term $\dot{H}/H^2$ from (\ref{H-dot-to-H2}) into
(\ref{q1}) yields
\begin{eqnarray}
\label{deceleration} q &=&
\frac{1}{2}(1-\Omega_{k}-3\Omega_{\Lambda}) +
\frac{1}{na}\Big({\frac{\Omega_{\Lambda}}{\gamma_n}}\Big)^{3/2}\Big(2\gamma_n
- \frac{G_{\rm c}}{G_{\rm g}}\nonumber\\&&-\frac{8\pi G_{\rm
c}\alpha H^2}{3n^4}\frac{\Omega_{\Lambda}}{\gamma_n}\Big).
\end{eqnarray}
Note that if we set $\alpha=\beta=0$ then from Eq.
(\ref{gamma-parameter1}) $\gamma_n=G_{\rm c}/G_{\rm g}$. Therefore
Eqs. (\ref{omegaD-eq-motion2}) and (\ref{deceleration}) reduce to
\begin{eqnarray}
{\Omega^{\prime}_{\Lambda}}= -\Omega_{\Lambda}
\Big[\Omega_k+(\Omega_{\Lambda}-1)\Big(3-\frac{2\Omega_{\Lambda}^{1/2}}{na}\Big(\frac{G_{\rm
g}}{G_{\rm c}}\Big)^{1/2}\Big)\Big],
\end{eqnarray}
\begin{equation}
q=\frac{1}{2}(1-\Omega_{k}-3\Omega_{\Lambda})+\frac{\Omega_{\Lambda}^{3/2}}{na}\Big(\frac{G_{\rm
g}}{G_{\rm c}}\Big)^{1/2},\label{q3}
\end{equation}
which are same as those results obtained for the NADE model in HL
cosmology (Jamil and Saridakis 2010).

Also for $G_{\rm c}\rightarrow G_{\rm g}$, Eqs.
(\ref{omegaD-eq-motion2}) and (\ref{deceleration}) recover the
results obtained for the ECNADE model in the standard FRW cosmology
(Karami and Sorouri 2010) as
\begin{eqnarray}
{\Omega^{\prime}_{\Lambda}}&=& \Omega_{\Lambda} \Big[3(1 -
\Omega_{\Lambda})-
\Omega_k+\frac{2}{na}\Big({\frac{\Omega_{\Lambda}}{\gamma_n}}\Big)^{1/2}
\nonumber\\&&\times\Big(\frac{\Omega_{\Lambda}-1}{\gamma_n}\Big)\Big(2\gamma_n
-1-\frac{\alpha
H^2}{3M_P^2{n^4}}\frac{\Omega_{\Lambda}}{\gamma_n}\Big)\Big],
\end{eqnarray}
\begin{eqnarray}
q &=& \frac{1}{2}(1-\Omega_{k}-3\Omega_{\Lambda})\nonumber\\&&+
\frac{1}{na}\Big({\frac{\Omega_{\Lambda}}{\gamma_n}}\Big)^{3/2}\Big(2\gamma_n
- 1-\frac{\alpha
H^2}{3{M^{2}_P}n^4}\frac{\Omega_{\Lambda}}{\gamma_n}\Big).
\end{eqnarray}
The EoS parameter of the ECNADE model can be parameterized as
(Huterer and Turner 1999, 2001; Weller and Albrecht 2001)
\begin{equation}
\omega_{\Lambda}(z)=\omega_0+\omega_1 z.\label{wpar}
\end{equation}
Using (\ref{eqpol}) and (\ref{wpar}), the ECNADE density evolves as
(Huterer and Turner 1999, 2001; Weller and Albrecht 2001; Copeland
et al. 2006)
\begin{equation}
\frac{\rho_{\Lambda}}{\rho_{\Lambda_{0}}}=a^{-3(1+\omega_{0}-\omega_{1})}e^{3\omega_{1}z}.\label{eqCDE}
\end{equation}
The Taylor expansion of the DE density around $a_0=1$ at the present
time yields
\begin{equation}
\ln{\rho_{\Lambda}}=\ln{\rho_{\Lambda_{0}}}+\frac{\rm
d{\ln{\rho_{\Lambda}}}}{{\rm d}
\ln{a}}\Big{|}_0\ln{a}+\frac{1}{2}\frac{\rm d^2
\ln{\rho_{\Lambda}}}{{\rm
d}({\ln{a}})^2}\Big{|}_0(\ln{a})^2+\cdots,\label{taylor expand}
\end{equation}
where the index 0 denotes the value of a quantity at present. Using
$\ln{a}=-\ln(1+z)\simeq -z+\frac{z^2}{2}$ for small redshifts, Eqs.
(\ref{eqCDE}) and (\ref{taylor expand}), respectively, reduce to
\begin{equation}
\frac{\ln{(\rho_{\Lambda}/\rho_{\Lambda_{0}})}}{\ln{a}}=-3(1+\omega_{0})-\frac{3}{2}\omega_{1}z\label{eqCDE1},
\end{equation}
\begin{equation}
\frac{\ln{(\rho_{\Lambda}/\rho_{\Lambda_{0}})}}{\ln{a}}=\frac{\rm
d{\ln{\rho_{\Lambda}}}}{{\rm d}
\ln{a}}\Big{|}_0-\frac{1}{2}\frac{\rm d^2 \ln{\rho_{\Lambda}}}{{\rm
d}({\ln{a}})^2}\Big{|}_0 z\label{eqCDE2}.
\end{equation}
Comparing Eqs. (\ref{eqCDE1}) and (\ref{eqCDE2}), one can obtain the
parameters $\omega_0$ and $\omega_1$ as
\begin{equation}
\omega_{0}=-\frac{1}{3}\frac{\rm d{\ln{\rho_{\Lambda}}}}{{\rm d}
\ln{a}}\Big{|}_0-1,\label{w0}
\end{equation}
\begin{equation}
\omega_{1}=\frac{1}{3}\frac{\rm d^2 \ln{\rho_{\Lambda}}}{{\rm d}
({\ln{a}})^2}\Big{|}_0.\label{w1}
\end{equation}
From Eq. (\ref{eqCDM}), the energy density of DM evolves as
$\rho_{\rm m}=\rho_{\rm m_{0}}a^{-3}$. Now using (\ref{eq10}) one
can get
\begin{equation}
\rho_{\Lambda}=\frac{\rho_{\rm m}}{\Omega_{\rm
m}}\Omega_{\Lambda}=\frac{\rho_{\rm
m_{0}}a^{-3}}{(1-\Omega_{k}-\Omega_{\Lambda})}\Omega_{\Lambda}.\label{rholambda}
\end{equation}
Substituting the above relation into (\ref{w0}) yields
\begin{equation}
\omega_{0}=-\frac{1}{3}\Big[\frac{\Omega^{\prime}_{\Lambda}}{\Omega_{\Lambda}}
+\frac{\Omega^{\prime}_{\Lambda}+\Omega^{\prime}_{k}}{(1-\Omega_{k}-\Omega_{\Lambda})}\Big]_0\label{eqomega0}.
\end{equation}
Adding Eqs. (\ref{omegaD-eq-motion2}) and (\ref{omega-prime-k}), we
get
\begin{eqnarray}
{\Omega^{\prime}_{\Lambda}}+{\Omega^{\prime}_{k}}&=&(1-\Omega_{k}-\Omega_{\Lambda})\Big[\Omega_{k}
+3\Omega_{\Lambda}-\frac{2}{na}
\Big({\frac{\Omega_{\Lambda}}{\gamma_n}}\Big)^{3/2}\nonumber\\&&\times\Big(2\gamma_n
-\frac{G_{\rm c}}{G_{\rm g}}-\frac{8\pi G_{\rm c}\alpha
H^2}{3{n^4}}\frac{\Omega_{\Lambda}}{\gamma_n}\Big)\Big].\label{omegalambdakpri}
\label{add-omega-prime}
\end{eqnarray}
Substituting Eqs. (\ref{omegaD-eq-motion2}) and
(\ref{add-omega-prime}) into (\ref{eqomega0}) yields
\begin{eqnarray}
\label{state-parameter} \omega_{0}&=&-1+
\frac{2}{3n\gamma_{n_0}}\Big({\frac{\Omega_{\Lambda_0}}{\gamma_{n_0}}}\Big)^{1/2}\Big(2\gamma_{n_0}
- \frac{G_{\rm c}}{G_{\rm g}}\nonumber\\&&-\frac{8\pi G_{\rm
c}\alpha
H_0^2}{3n^4}\frac{\Omega_{\Lambda_0}}{\gamma_{n_0}}\Big).\label{omega0-condition}
\end{eqnarray}
Note that in the absence of correction terms, i.e. $\alpha=\beta=0$
and $\gamma_n=G_{\rm c}/G_{\rm g}$, Eq. (\ref{state-parameter})
yields
\begin{eqnarray}
\omega_0=-1+\frac{2\Omega_{\Lambda_0}^{1/2}}{3{ n}}\Big(\frac{G_{\rm
g}}{G_{\rm c}}\Big)^{1/2},
\end{eqnarray}
which is the EoS parameter of the NADE model in HL cosmology (Jamil
and Saridakis 2010).

Also in the limit of $G_{\rm c}\rightarrow G_{\rm g}$, Eq.
(\ref{state-parameter}) recovers the EoS parameter of the ECNADE
model in the standard FRW cosmology (Karami and Sorouri 2010) as
\begin{eqnarray}
\omega_0= -1 +
\frac{2}{3n\gamma_{n_0}}\Big({\frac{\Omega_{\Lambda_0}}{\gamma_{n_0}}}\Big)^{1/2}
~~~~~~~~~~~~~\nonumber\\\times\Big(2\gamma_{n_0} - 1-\frac{\alpha
H_0^2}{3{M^{2}_P}n^4}\frac{\Omega_{\Lambda_0}}{\gamma_{n_0}}\Big).
\end{eqnarray}
Using Eqs. (\ref{w1}) and (\ref{rholambda}), one can get
\begin{eqnarray}
\omega_{1}=\frac{1}{3}\Big[\frac{{\Omega}^{\prime\prime}_{\Lambda}}{\Omega_{\Lambda}}
-\frac{{\Omega}^{\prime2}_{\Lambda}}{\Omega_{\Lambda}^2}
+\frac{\Omega^{\prime\prime}_{\Lambda}+{\Omega}^{\prime\prime}_{k}}{(1-\Omega_{k}-\Omega_{\Lambda})}
\nonumber\\+\frac{(\Omega^{\prime}_{\Lambda}+\Omega^{\prime}_{k})^2}{(1-\Omega_{k}-\Omega_{\Lambda})^2}\Big]_0.\label{w1-2}
\end{eqnarray}
Taking derivative of Eqs. (\ref{omegaD-eq-motion2}) and
(\ref{omega-prime-k}) with respect to $x=\ln{a}$, one can obtain
\begin{eqnarray}
\Omega^{\prime\prime}_{\Lambda}&=&-(\Omega_{\Lambda}\Omega^{\prime}_{k}+\Omega^{\prime}_
{\Lambda}\Omega_{k})
+3\Omega^{\prime}_{\Lambda}(1-2\Omega_{\Lambda})\nonumber\\&&+(\Omega_{\Lambda}-\Omega^{\prime}_
{\Lambda}-1)\Big[\frac{G_{\rm c}}{G_{\rm
g}}\frac{2n^2}{aH^3\eta^3}-\frac{4\Omega_{\Lambda}}{aH\eta}+\frac{16\pi{G_{\rm
c}}\alpha}{3aH^3\eta^5}\Big]\nonumber\\&&+(1-\Omega_{\Lambda})
\Big[\Big(\frac{4\Omega_{\Lambda}}{aH\eta}-\frac{G_{\rm c}}{G_{\rm
g}}\frac{6n^2}{aH^3\eta^3}\Big)\Big(\frac{\dot{H}}{H^2}+\frac{\dot{\eta}}{H\eta}\Big)
\nonumber\\&&-\frac{4\Omega^{\prime}_{\Lambda}}{aH\eta}-\frac{16\pi{G_{\rm
c}}\alpha}{3aH^3\eta^5}\Big(\frac{3\dot{H}}{H^2}+\frac{5\dot{\eta}}{H\eta}\Big)\Big],
\end{eqnarray}
and
\begin{eqnarray}
\Omega^{\prime\prime}_{k}&=&\Omega^{\prime}_{k}(1-2\Omega_{k})-3(\Omega_{\Lambda}\Omega^
{\prime}_{k}+\Omega^{\prime}_{\Lambda}\Omega_{k})
+(\Omega_{k}-\Omega^{\prime}_{k})\nonumber\\&&\times
\Big[\frac{G_{\rm c}}{G_{\rm
g}}\frac{2n^2}{aH^3\eta^3}-\frac{4\Omega_{\Lambda}}{aH\eta}+\frac{16\pi{G_{\rm
c}}\alpha}{3aH^3\eta^5}\Big]\nonumber\\&&-
\Omega_{k}\Big[\Big(\frac{4\Omega_{\Lambda}}{aH\eta}-\frac{G_{\rm
c}}{G_{\rm
g}}\frac{6n^2}{aH^3\eta^3}\Big)\Big(\frac{\dot{H}}{H^2}+\frac{\dot{\eta}}{H\eta}\Big)
\nonumber\\&&-\frac{4\Omega^{\prime}_{\Lambda}}{aH\eta}-\frac{16\pi{G_{\rm
c}}\alpha}{3aH^3\eta^5}\Big(\frac{3\dot{H}}{H^2}+\frac{5\dot{\eta}}{H\eta}\Big)\Big].
\end{eqnarray}
Using Eq. (\ref{state-parameter}), the above expressions for
$\Omega^{\prime\prime}_{\Lambda}$ and $\Omega^{\prime\prime}_{k}$
can be rewritten as
\begin{eqnarray}
\Omega^{\prime\prime}_{\Lambda}&=&-(\Omega_{\Lambda}\Omega^{\prime}_{k}+\Omega^{\prime}_
{\Lambda}\Omega_{k})
-3\Omega_{\Lambda}(\Omega_{\Lambda}-\Omega^{\prime}_{\Lambda}-1)\omega_{0}
\nonumber\\&&+(1-\Omega_{\Lambda})\Big[3(\Omega_{\Lambda}+\Omega^{\prime}_{\Lambda})+A\Big],\label{omegalambdapri}
\end{eqnarray}
\begin{eqnarray}
\Omega^{\prime\prime}_{k}&=&\Omega^{\prime}_{k}(1-2\Omega_{k})
-3\Omega_{\Lambda}(\Omega_{k}-\Omega^{\prime}_{k})\omega_{0}
\nonumber\\&&-\Omega_{k}\Big[3(\Omega_{\Lambda}+\Omega^{\prime}_{\Lambda})+A\Big],\label{omegakpri}
\end{eqnarray}
where
\begin{eqnarray}
A&=&\Big(\frac{4\Omega_{\Lambda}}{aH\eta}-\frac{G_{\rm c}}{G_{\rm
g}}\frac{6n^2}{aH^3\eta^3}\Big)\Big(\frac{\dot{H}}{H^2}+\frac{\dot{\eta}}{H\eta}\Big)
-\frac{4\Omega^{\prime}_{\Lambda}}{aH\eta}\nonumber\\&&-\frac{16\pi{G_{\rm
c}}\alpha}{3aH^3\eta^5}\Big(\frac{3\dot{H}}{H^2}+\frac{5\dot{\eta}}{H\eta}\Big).\label{A}
\end{eqnarray}
Using Eqs. (\ref{density-nade-omega}), (\ref{q1}) and
$\dot{\eta}=1/a$, one can rewrite (\ref{A}) as
\begin{eqnarray}
A&=&-9\Omega_{\Lambda}(1+q)(1+\omega_{0})\nonumber
\\&&+\frac{\Omega_{\Lambda}}{na}
\Big(\frac{\Omega_{\Lambda}}{\gamma_{n}}\Big)^{1/2}
\Big[17+4(\Omega_{k}+2q)\nonumber
\\&&+3(7-4\Omega_{\Lambda})\omega_{0}
\Big.-\frac{8}{na}\Big(\frac{\Omega_{\Lambda}}{\gamma_{n}}\Big)^{1/2}
\nonumber\\&&-\frac{32\pi G_{\rm c}\alpha
H^2}{3an^5\Omega_{\Lambda}}
\Big(\frac{\Omega_{\Lambda}}{\gamma_{n}}\Big)^{5/2}\Big].\label{A2}
\end{eqnarray}
Adding Eqs. (\ref{omegalambdapri}) and (\ref{omegakpri}), we get
\begin{eqnarray}
\Omega^{\prime\prime}_{\Lambda}+\Omega^{\prime\prime}_{k}&=&-(\Omega_{\Lambda}\Omega^
{\prime}_{k}+\Omega^{\prime}_{\Lambda}\Omega_{k})
+\Omega^{\prime}_{k}(1-2\Omega_{k})\nonumber\\&&-3(\Omega_{k}-\Omega^{\prime}_{k}+\Omega_
{\Lambda}-\Omega^{\prime}_{\Lambda}-1)\Omega_{\Lambda}\omega_0
\nonumber\\&&+(1-\Omega_{\Lambda}-\Omega_{k})\Big[3(\Omega_{\Lambda}+\Omega^{\prime}_{\Lambda})+A\Big].\label{omegalambdakppri}
\end{eqnarray}
Substituting Eqs. (\ref{omegaD-eq-motion2}),
(\ref{omegalambdakpri}), (\ref{omegalambdapri}) and
(\ref{omegalambdakppri}) into (\ref{w1-2}) yields
\begin{equation}
\omega_{1}=(1+\omega_{0})\Big[1-\Omega_{k_0}-3(1-\Omega_{\Lambda_0})\omega_{0}\Big]
+\frac{A_0}{3\Omega_{\Lambda_0}},
\end{equation}
which can be rewritten by the help of Eq. (\ref{A2}) as
\begin{eqnarray}
\omega_{1}&=&(1+\omega_{0})\Big[1-\Omega_{k_0}-3(1-\Omega_{\Lambda_0})\omega_{0}\Big]
\nonumber\\&&-3(1+q_0)(1+\omega_{0})+\frac{1}{3n}
\Big(\frac{\Omega_{\Lambda_0}}{\gamma_{n_0}}\Big)^{1/2}
\nonumber\\&&\times\Big[17+4(\Omega_{k_0}+2q_0)+3(7-4\Omega_{\Lambda_0})\omega_{0}\nonumber
\\&&-\Big.\frac{8}{n}\Big(\frac{\Omega_{\Lambda_0}}{\gamma_{n_0}}\Big)^{1/2}
-\frac{32\pi G_{\rm c}\alpha H_0^2}{3n^5\Omega_{\Lambda_0}}
\Big(\frac{\Omega_{\Lambda_0}}{\gamma_{n_0}}\Big)^{5/2}\Big].\label{EoS1}
\end{eqnarray}
\subsection{Interacting case}
Here, we consider a case in which the ECNADE and DM interact with
each other. This causes the energy conservation law for each dark
component not to be held separately, i.e.
\begin{equation}
\dot{\rho}_{\Lambda}+3H(1+\omega_{\Lambda})\rho_{\Lambda}=-Q,\label{eqintDE}
\end{equation}
\begin{equation}
\dot{\rho}_{\rm m}+3H\rho_{\rm m}=Q,\label{eqintCDM}
\end{equation}
where $Q=3b^2H\rho_{\Lambda}$ (Pav\'{o}n and Zimdahl 2005) stands
for the interaction term with coupling constant $b^2$. For $Q>0$,
there is an energy transfer from ECNADE to DM. The recent
observational evidence provided by the galaxy cluster Abell A586
supports the interaction between DE and DM (Bertolami et al. 2007,
2009; Abdalla et al. 2009).

Taking time derivative of the first Friedmann Eq. (\ref{Fr1fluid})
and using (\ref{eqomega}), (\ref{eq10}),
 (\ref{density-nade-omega}),
(\ref{gamma-parameter1}), (\ref{rhodot}) and (\ref{eqintCDM}) yields
\begin{eqnarray}
\frac{\dot{H}}{H^2} &=&\frac{1}{2}\Big[\Omega_k-3(1 -
\Omega_{\Lambda})+3b^2\Omega_{\Lambda}\Big]
-\frac{1}{na}\Big({\frac{\Omega_{\Lambda}}{\gamma_n}}\Big)^{3/2}\nonumber\\&&\times\Big(2\gamma_n
- \frac{G_{\rm c}}{G_{\rm g}}-\frac{8\pi G_{\rm c}\alpha
H^2}{3n^4}\frac{\Omega_{\Lambda}}{\gamma_n}\Big).\label{H-dot-to-H2-interact}
\end{eqnarray}
Substituting this into Eqs. (\ref{omegaD-eq-motion1}) and
(\ref{omegak-eq-motion1}) one can get
\begin{eqnarray}
{\Omega^{\prime}_{\Lambda}}= \Omega_{\Lambda} \Big[3(1 -
\Omega_{\Lambda})-3b^2{\Omega}_{\Lambda}
-\Omega_k+\frac{2}{na}\Big({\frac{\Omega_{\Lambda}}{\gamma_n}}\Big)^{1/2}
\nonumber\\\times\Big(\frac{\Omega_{\Lambda}-1}{\gamma_n}\Big)\Big(2\gamma_n
-\frac{G_{\rm c}}{G_{\rm g}}-\frac{8\pi G_{\rm c}\alpha
H^2}{3{n^4}}\frac{\Omega_{\Lambda}}{\gamma_n}\Big)\Big],\label{omegaD-eq-motion3}
\end{eqnarray}
and
\begin{eqnarray}
{\Omega^{\prime}_{k}}&=&\Omega_{k}\Big[(1
-\Omega_k)-3\Omega_{\Lambda}-3b^2\Omega_{\Lambda}
+\frac{2}{na}\Big({\frac{\Omega_{\Lambda}}{\gamma_n}}\Big)^{3/2}\nonumber\\&&\times\Big(2\gamma_n
-\frac{G_{\rm c}}{G_{\rm g}}-\frac{8\pi G_{\rm c}\alpha
H^2}{3{n^4}}\frac{\Omega_{\Lambda}}{\gamma_n}\Big)\Big].\label{omega-prime-k-interact}
\end{eqnarray}
Replacing the term $\dot{H}/H^2$ from (\ref{H-dot-to-H2-interact})
into (\ref{q1}) yields
\begin{eqnarray}
\label{deceleration-int} q &=&
\frac{1}{2}(1-\Omega_{k}-3\Omega_{\Lambda}) +
\frac{1}{na}\Big({\frac{\Omega_{\Lambda}}{\gamma_n}}\Big)^{3/2}\Big(2\gamma_n
\nonumber\\&&- \frac{G_{\rm c}}{G_{\rm g}}-\frac{8\pi G_{\rm
c}\alpha
H^2}{3n^4}\frac{\Omega_{\Lambda}}{\gamma_n}\Big)-\frac{3}{2}b^2\Omega_{\Lambda}.
\end{eqnarray}
From
\begin{equation}
\rho_{\Lambda}=\frac{\rho_{\rm m}}{\Omega_{\rm
m}}\Omega_{\Lambda}=\frac{\rho_{\rm
m}}{(1-\Omega_{k}-\Omega_{\Lambda})}\Omega_{\Lambda},\label{rholambdaint}
\end{equation}
one can obtain
\begin{equation}
\frac{\rm d{\ln{\rho_{\Lambda}}}}{{\rm d}
\ln{a}}=\frac{\rho^{\prime}_{\rm m}}{\rho_{\rm
m}}-\frac{\Omega^{\prime}_{\rm m}}{\Omega_{\rm
m}}+\frac{\Omega^{\prime}_{\Lambda}}{\Omega_{\Lambda}}.\label{drholambda}
\end{equation}
From Eq. (\ref{eqintDE}) and using (\ref{wpar}), the interacting
ECNADE density evolves as
\begin{equation}
\frac{\rho_{\Lambda}}{\rho_{\Lambda_{0}}}=a^{-3(1+\omega_{0}-\omega_{1}+b^2)}e^{3\omega_{1}z}.\label{eqCDEint}
\end{equation}
Using $\ln{a}=-\ln(1+z)\simeq -z+\frac{z^2}{2}$ for small redshifts,
Eq. (\ref{eqCDEint}) reduces to
\begin{equation}
\frac{\ln{(\rho_{\Lambda}/\rho_{\Lambda_{0}})}}{\ln{a}}=-3(1+\omega_{0}+b^2)-\frac{3}{2}\omega_1
z.\label{eqCDE1int}
\end{equation}
Comparing Eqs. (\ref{eqCDE2}) and (\ref{eqCDE1int}), one can obtain
the parameters $\omega_0$ and $\omega_1$ for the interacting case as
\begin{equation}
\omega_{0}=-\frac{1}{3}\frac{\rm d{\ln{\rho_{\Lambda}}}}{{\rm d}
\ln{a}}{\Big |}_0-1-b^2,\label{w0int}
\end{equation}
\begin{equation}
\omega_{1}=\frac{1}{3}\frac{\rm d^2 \ln{\rho_{\Lambda}}}{{\rm d}
({\ln{a}})^2}{\Big |}_0.\label{w1int}
\end{equation}
Substituting Eq. (\ref{drholambda}) into (\ref{w0int}) and using
(\ref{eqintCDM}) one can get
\begin{eqnarray}
\omega_{0}=-\frac{1}{3}\Big[\frac{\Omega^{\prime}_{\Lambda}}{\Omega_{\Lambda}}
+\frac{\Omega^{\prime}_{\Lambda}+\Omega^{\prime}_{k}}{(1-\Omega_{k}-\Omega_{\Lambda})}\Big]_0
\nonumber\\-b^2\Big(\frac{1-\Omega_{k}}{1-\Omega_{k}-\Omega_{\Lambda}}\Big)_0.
\label{eqomega0-interact}
\end{eqnarray}
Using Eqs. (\ref{rholambdaint}) and (\ref{w1int}) one can obtain
\begin{eqnarray}
\omega_{1}=\frac{1}{3}\Big[\frac{3b^2\Omega^{\prime}_{\Lambda}}{1-\Omega_{k}-\Omega_{\Lambda}}+\frac{3b^2\Omega_{\Lambda}(\Omega^{\prime}_{\Lambda}+\Omega^{\prime}_{k})}{(1-\Omega_{k}-\Omega_{\Lambda})^2}+\frac{{\Omega}^{\prime\prime}_{\Lambda}}{\Omega_{\Lambda}}
\nonumber\\-\frac{{\Omega}^{\prime2}_{\Lambda}}{\Omega_{\Lambda}^2}
+\frac{\Omega^{\prime\prime}_{\Lambda}+{\Omega}^{\prime\prime}_{k}}{(1-\Omega_{k}-\Omega_{\Lambda})}
+\frac{(\Omega^{\prime}_{\Lambda}+\Omega^{\prime}_{k})^2}{(1-\Omega_{k}-\Omega_{\Lambda})^2}\Big]_0.\label{w1-2int}
\end{eqnarray}
Adding Eqs. (\ref{omegaD-eq-motion3}) and
(\ref{omega-prime-k-interact}) gives
\begin{eqnarray}
{\Omega^{\prime}_{\Lambda}}+{\Omega^{\prime}_{k}}&=&(1-\Omega_{k}-\Omega_{\Lambda})
\Big[\Omega_{k}+3\Omega_{\Lambda}-\frac{2}{na}
\Big({\frac{\Omega_{\Lambda}}{\gamma_n}}\Big)^{3/2}\nonumber\\&&\times\Big(2\gamma_n
-\frac{G_{\rm c}}{G_{\rm g}}-\frac{8\pi G_{\rm c}\alpha
H^2}{3{n^4}}\frac{\Omega_{\Lambda}}{\gamma_n}\Big)\Big]\nonumber\\
&&-3b^2\Omega_{\Lambda}(\Omega_{k}+\Omega_{\Lambda}).
\label{add-omega-prime-interact}
\end{eqnarray}
Substituting Eqs. (\ref{omegaD-eq-motion3}) and
(\ref{add-omega-prime-interact}) into (\ref{eqomega0-interact})
yields
\begin{eqnarray}
\label{state-parameter-2-interact} \omega_{0}=-1+
\frac{2}{3n\gamma_{n_0}}\Big({\frac{\Omega_{\Lambda_0}}{\gamma_{n_0}}}\Big)^{1/2}
~~~~~~~~~~~~~~~~~~~~~~~~~~\nonumber\\\times\Big(2\gamma_{n_0} -
\frac{G_{\rm c}}{G_{\rm g}}-\frac{8\pi G_{\rm c}\alpha
H_0^2}{3n^4}\frac{\Omega_{\Lambda_0}}{\gamma_{n_0}}\Big)-b^2,
\end{eqnarray}
which in the absence of interaction, i.e. $b^2=0$, reduces to Eq.
(\ref{state-parameter}). The EoS parameter $\omega_0$ versus the
interacting coupling parameter $b^2$ is plotted in Fig. \ref{w0-b2}.
Figure shows that in the absence of interaction ($b^2=0$),
$\omega_0=-0.85$ and cannot cross the phantom divide line. However,
in the presence of interaction, $b^2\neq 0$, the phantom EoS
$\omega_0<-1$ can be obtained when $b^2\geq0.15$ for the coupling
between ECNADE and DM. This value for coupling constant $b^2$ is
consistent with recent observations (Wang et al. 2005a, 2005b). Also
the phantom divide crossing is compatible with the observation
(Komatsu et al. 2011).

Taking derivative of Eqs. (\ref{omegaD-eq-motion3}) and
(\ref{omega-prime-k-interact}) with respect to $x=\ln a$ and using
(\ref{state-parameter-2-interact}), one can get
\begin{eqnarray}
\Omega^{\prime\prime}_{\Lambda}&=&-(\Omega_{\Lambda}\Omega^{\prime}_{k}+\Omega^{\prime}_{\Lambda}\Omega_{k})
\nonumber\\&&-3\Omega_{\Lambda}(\Omega_{\Lambda}-\Omega^{\prime}_{\Lambda}-1)(\omega_{0}+b^2)
\nonumber\\&&+(1-\Omega_{\Lambda})\Big[3(\Omega_{\Lambda}+\Omega^{\prime}_{\Lambda})+A\Big]
\nonumber\\&&-6b^2\Omega^{\prime}_{\Lambda}\Omega_{\Lambda},\label{omegalambdapri-interact}
\end{eqnarray}
and
\begin{eqnarray}
\Omega^{\prime\prime}_{k}&=&\Omega^{\prime}_{k}(1-2\Omega_{k})
-3\Omega_{\Lambda}(\Omega_{k}-\Omega^{\prime}_{k})(\omega_{0}+b^2)
\nonumber\\&&-\Omega_{k}\Big[3(\Omega_{\Lambda}+\Omega^{\prime}_{\Lambda})+A\Big]
\nonumber\\&&-3b^2(\Omega^{\prime}_{\rm
k}\Omega_{\Lambda}+\Omega_{\rm
k}\Omega^{\prime}_{\Lambda}),\label{omegakpri-interact}
\end{eqnarray}
where $A$ is given by Eq. (\ref{A}). Using Eqs.
(\ref{density-nade-omega}), (\ref{q1}),
(\ref{H-dot-to-H2-interact}), (\ref{state-parameter-2-interact}) and
$\dot{\eta}=1/a$, one can rewrite (\ref{A}) for the interacting case
as
\begin{eqnarray}
A&=&-9\Omega_{\Lambda}(1+q)(1+\omega_{0}+b^2)+\frac{\Omega_{\Lambda}}{na}
\Big(\frac{\Omega_{\Lambda}}{\gamma_{n}}\Big)^{1/2}
\nonumber\\&&\times\Big[17+4(\Omega_{k}+2q)+3(7-4\Omega_{\Lambda})\omega_{0}+21b^2
\nonumber
\\&&\Big.-\frac{8}{na}\Big(\frac{\Omega_{\Lambda}}{\gamma_{n}}\Big)^{1/2}
-\frac{32\pi G_{\rm c}\alpha H^2}{3an^5\Omega_{\Lambda}}
\Big(\frac{\Omega_{\Lambda}}{\gamma_{n}}\Big)^{5/2}\Big].\label{A2-int}
\end{eqnarray}
Adding Eqs. (\ref{omegalambdapri-interact}) and
(\ref{omegakpri-interact}) gives
\begin{eqnarray}
\Omega^{\prime\prime}_{\Lambda}+\Omega^{\prime\prime}_{k}&=&-(\Omega_{\Lambda}\Omega^{\prime}_{k}
+\Omega^{\prime}_{\Lambda}\Omega_{k})
+\Omega^{\prime}_{k}(1-2\Omega_{k})\nonumber\\&&-3\Omega_{\Lambda}(\omega_0+b^2)(\Omega_{k}
-\Omega^{\prime}_{k}+\Omega_{\Lambda}-\Omega^{\prime}_{\Lambda}-1)
\nonumber\\&&+(1-\Omega_{\Lambda}-\Omega_{k})\Big[3(\Omega_{\Lambda}+\Omega^{\prime}
_{\Lambda})+A\Big]\nonumber\\&& -3b^2(\Omega^{\prime}_{\rm
k}\Omega_{\Lambda}+\Omega_{\rm
k}\Omega^{\prime}_{\Lambda})-6b^2\Omega^{\prime}_{\Lambda}\Omega_{\Lambda}.\label{omegalambdakppri-interact}
\end{eqnarray}
Replacing Eqs. (\ref{omegaD-eq-motion3}),
(\ref{add-omega-prime-interact}), (\ref{omegalambdapri-interact})
and (\ref{omegalambdakppri-interact}) into (\ref{w1-2int}) gives
\begin{eqnarray}
\omega_{1}&=&\frac{b^2\Omega_{\Lambda_0}}{(1-\Omega_{\Lambda_0}-\Omega_{k_0})}
\Big[-3(\omega_{0}+b^2)\nonumber\\&&+6\Omega_{\Lambda_0}\omega_{0}-2\Omega_{k_0}\Big]
\nonumber\\&&+(3\Omega_{\Lambda_0}\omega_{0}-\Omega_{k_0})
\Big[\frac{\Omega_{k_0}}{3}-\Omega_{\Lambda_0}\omega_{0}\nonumber\\&&-\frac{b^2\Omega_
{\Lambda_0}}{(1-\Omega_{\Lambda_0}-\Omega_{k_0})}\Big]
\nonumber\\
&&+\Big[\frac{b^2\Omega_{\Lambda_0}}{(1-\Omega_{\Lambda_0}-\Omega_{k_0})}-
\Omega_{\Lambda_0}\omega_{0}+\frac{\Omega_{k_0}}{3}-1\Big]\nonumber\\&&
\times\Big[3\omega_{0}(1-\Omega_{\Lambda_0})+3b^2+\Omega_{k_0}\Big]\nonumber\\&&+
(\omega_{0}+b^2)
\Big[3(\omega_{0}+b^2)(2\Omega_{\Lambda_0}-1)\nonumber\\&&+1-2\Omega_{k_0}-6b^2\Omega_{
\Lambda_0}\Big]+\frac{A_0}{3\Omega_{\Lambda_0}}+1, \label{EoS1-int}
\end{eqnarray}
which in the absence of interaction, i.e. $b^2=0$, yields Eq.
(\ref{EoS1}). The EoS parameter $\omega_1$ versus the interacting
coupling parameter $b^2$ is plotted in Fig. \ref{w1-b2}. Figure
shows that in the absence of interaction ($b^2=0$), $\omega_1=0.07$
and the EoS parameter $\omega_1$ increases when the interacting
coupling parameter $b^2$ increases. This is in agreement with the
result obtained by Jamil and Saridakis (2010).
\section{Conclusions}

Here, we investigated the ECNADE scenario in a FRW Universe in the
framework of HL gravity. We considered an arbitrary spatial local
curvature for the background geometry and allowed for an interaction
between the ECNADE and DM. In both the regular and interacting case
we obtained the deceleration parameter as well as the differential
equations which determine the evolution of the ECNADE density
parameter. Finally, using a low redshift expansion of the EoS
parameter of DE as
 $\omega_{\Lambda}(z)=\omega_0 +\omega_1
z$, we calculated $\omega_0$ and $\omega_1$ as functions of the DE
and curvature density parameters, $\Omega_{\Lambda_0}$ and
$\Omega_{k_0}$ respectively, of the running parameter
$\lambda=(2\frac{G_{\rm g}}{G_{\rm c}}+1)/3$ of HL gravity, of the
parameter $n$ of ECNADE, of the interaction coupling $b^2$, and of
the coefficients of correction terms $\alpha$ and $\beta$.

Interestingly enough we found that in the presence of interaction
between ECNADE and DM, the EoS parameter $\omega_0$ of ECNADE in HL
gravity, can cross the phantom divide line which is compatible with
the recent observations. The interaction between ECNADE and DM can
be detected in the formation of large scale structures. It was
suggested that the dynamical equilibrium of collapsed structures
such as galaxy clusters would be modified due to the coupling
between DE and DM (Bertolami et al. 2007, 2009; Abdalla et al.
2009). The idea is that the virial theorem is modified by the energy
exchange between the dark sectors leading to a bias in the
estimation of the virial masses of clusters when the usual virial
conditions are employed. This provides a probe in the near Universe
of the dark coupling. The other observational signatures on the dark
sectors' mutual interaction can be found in the probes of the cosmic
expansion history by using the type Ia supernovae (SNeIa), baryonic
acoustic oscillation (BAO) and cosmic microwave background (CMB)
shift data (Guo and Ohta 2007; Feng et al. 2008; He et al. 2009;
Honorez et al. 2010).
\\
\\
\noindent{\textbf{Acknowledgements}\\ The work of K. Karami has been
supported financially by Research Institute for Astronomy and
Astrophysics of Maragha (RIAAM) under research project No. 1/2343.
We would like to thank the anonymous referee for giving useful and
enlightening comments to improve this paper.

\begin{figure}\centering
\includegraphics[scale=.83]{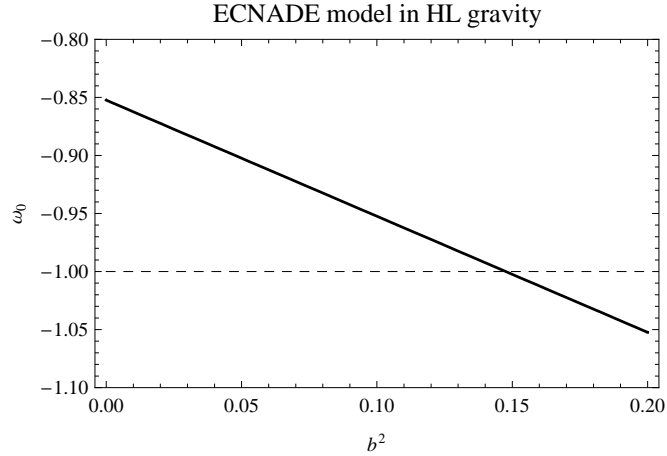}\\
\caption{The EoS parameter $\omega_0$ of the ECNADE in HL gravity,
Eq. (\ref{state-parameter-2-interact}), versus the interacting
coupling parameter $b^2$. Auxiliary parameters are: $n=2.716$ (Wei
and Cai 2008b), $\alpha=-7.5$, $\gamma_{n_0}=15$ (Karami and
Abdolmaleki 2010c), $\Omega_{\Lambda_0}=0.728$,
$\Omega_{k_0}=-0.013$ (Komatsu et al. 2011), $\lambda=1.02$ (Dutta
and Saridakis 2010), $G_{\rm c}/G_{\rm g}=2/(3\lambda-1)$,
$H_0=74.2~{\rm Km~S^{-1}~Mpc^{-1}}$ (Riess et al. 2009) and
$M_P^{-2}=8\pi G_{\rm g}=1$.}   \label{w0-b2}
\end{figure}

\begin{figure}\centering
\includegraphics[scale=.83]{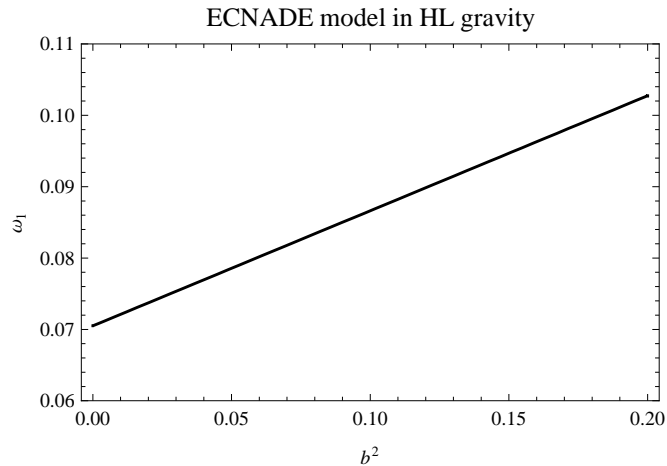}\\
\caption{The EoS parameter $\omega_1$ of the ECNADE in HL gravity,
Eq. (\ref{EoS1-int}), versus the interacting coupling parameter
$b^2$. Auxiliary parameters as in Fig. \ref{w0-b2}.}   \label{w1-b2}
\end{figure}


\end{document}